\documentclass[12pt,twoside]{article}
\usepackage{fleqn,epsfig,espcrc1}
\usepackage{graphicx}

\def\ubar#1{\overline{u}(#1)}
\def\u#1{u(#1)}

\def\ks {\not\!k}
\def\xs {\not\!x}

\begin{document}

\title{\Large
 \bf Determination of the $\Delta^{++}$ magnetic dipole moment}

\vskip 0.5cm

\author{\large G. L\'opez Castro \address[CIN]{Departamento
F\'\i sica, Centro de
Investigaci\'on
y de Estudios Avanzados del IPN, \\ Apdo. Postal 14-740, 07000
M\'exico, D.F., M\'exico}
and A. Mariano
\address[UNLP]{ Departamento de F\'\i sica, Facultad de
Ciencias
Exactas, Universidad Nacional de La Plata,
cc. 67, 1900 La Plata, Argentina}}

\maketitle

\begin{abstract}

We study the elastic and radiative $\pi^+p$ scattering within a full
dynamical model which incorporates the finite width effects of the
$\Delta^{++}$.
The scattering amplitudes are invariant under contact transformations of
the spin 3/2 field and gauge-invariance is fulfilled
for the radiative case. The pole parameters of the $\Delta^{++}$ obtained
from the elastic cross section are $m_{\Delta} = (1211.2 \pm 0.4)$ MeV
and $\Gamma_{\Delta} = (88.2 \pm 0.4)$ MeV. From a fit to the most
sensitive observables in radiative $\pi^+p$ scattering, we obtain
$\mu_{\Delta} = (6.14 \pm 0.51)e/2m_p$ for the magnetic dipole moment of
the $\Delta^{++}$.

\end{abstract}
pacs: 13.75.Gx, 14.20.Gk, 13.40.Em

 \vskip 1.cm

 The description of resonances in particle physics has gotten a renewed
interest with the advent of precise measurements of the $Z^0$ gauge boson
properties at LEP \cite{pdg98}. The idea behind these recent works is to
provide a consistent description of
resonances based on general principles of quantum field theory such as
gauge invariance and analyticity \cite{z0}.
 On another hand, it has long been recognized that mass and width are
physical properties of resonances that can be
determined in a model-independent and gauge-invariant way by identifying
the pole position of the S-matrix amplitude \cite{smatrix}.

However, the determination of the couplings of resonances to other
particles necessarily involves the assumption of  a dynamical model to
describe  how they enter the relevant amplitude.
 It is not idle to mention that most of the values of masses, widths and
branching ratios for hadronic resonances quoted in the Particle Data
Book \cite{pdg98} correspond to parameters obtained from somewhat
arbitrary parametrizations of the Breit-Wigner formula.

  The aim of the present paper is to determine the magnetic dipole moment
(MDM) of the $\Delta^{++}$ resonance by using a {\it full} dynamical model
which consistently describes elastic and radiative $\pi^+p$ scattering
data.  The model, to be described below, reproduces very well the
total and differential cross sections for elastic $\pi^+p$ scattering
close to the resonance
region. This model also provides an amplitude for radiative $\pi^+p$
scattering that satisfies electromagnetic gauge invariance when finite
width effects of the $\Delta^{++}$ resonance are taken into account.

The spirit of our calculations is similar to the approaches developed in
Refs. \cite{fls} to cure gauge invariance problems associated to a naive
introduction of the finite width of the resonance.
As it has been shown for the case of the $W^{\pm}$ gauge boson,
the use of  propagators and electromagnetic vertices that include
absorptive corrections coming from loops of fermions allows to introduce
finite width effects in scattering amplitudes in a gauge-invariant way
\cite{fls}. Similar conclusions has been reached for the $\rho^{\pm}$
unstable meson by including loops of pions in absorptive corrections
to its  propagator and electromagnetic vertex \cite{bls}. The
expressions for the propagator and electromagnetic vertex of unstable
particles, when massless particles appears in absorptive loop corrections,
are equivalent to the ones obtained by using a {\it complex mass scheme}.
In the complex mass scheme, gauge invariance of the amplitudes is
satisfied
if the squared mass $M^2$ of unstable particles in  Feynman rules is
replaced by $M^2 -iM\Gamma$ with $\Gamma$ being the decay width
\cite{w}.

It is interesting to note that the mass and width parameters of the
$\Delta^{++}$ resonance
required to describe the $\pi^+p$ elastic scattering within our model,
are consistent with the ones obtained from a model-independent
analysis of this process \cite{bernicha}. This feature cast confidence
on the consistency of the dynamical model advocated in this paper. Thus,
the magnetic dipole moment of the $\Delta^{++}$ turns out to be the only
adjustable parameter required to describe radiative $\pi^+p$
scattering.

  Some of the previous determinations of the $\Delta^{++}$ MDM have been
summarized in Ref. \cite{pdg98}. Due to the large spread of
central values, the Particle Data Group \cite{pdg98} prefers to quote a
rough estimate for this multipole which lie in the range $\mu_{\Delta}
\sim 3.7$ to $7.5$ in  units of $e/2m_p$.
The most recent determinations of the $\Delta^{++}$ MDM are based on fits
to the radiative $\pi^+p$ scattering data
of the SIN \cite{sin} and UCLA \cite{ucla} experiments. Some of the models
\cite{Lin}  used to extract the MDM rely on the soft photon theorem
\cite{low}, and on a specific parametrization of the
off-shell elastic amplitude to fix the terms of order $\omega_{\gamma}^0$
($\omega_{\gamma}$ is the photon energy in the radiative process) by
requiring gauge-invariance. Furthermore, Ref. \cite{Lin} ignores the
effects of the finite width of the $\Delta^{++}$ and diagrams with
vertices involving four particles (see Figs. 1(e-f) in Ref.
\cite{elamiri}).

Invariance under contact transformations ensures that physical
amplitudes involving the $\Delta$ resonance are independent of any
arbitrariness in the Feynman rules of a given theoretical model
for this resonance\cite{surdarshan,etemadi}. Vertices and
propagators depend on an arbitrary parameter $A$ that changes as
$A \rightarrow A' = (A -2a)/(1+4a)$, when the transformation $
\psi^{\mu} \rightarrow \psi^{\mu} +
a\gamma^{\mu}\gamma_{\alpha}\psi^{\alpha}$($a\not = -1/4$) is
done. Physical amplitudes, should however be independent of $A$.  Other
models (see for example \cite{wittman}) make use
of an amplitude that depends on $A$ ;hence, the value of the MDM
is quoted for an specific value of this arbitrary parameter. In
Ref. \cite{pt} a determination of the MDM free of ambiguities
related to contact transformations is provided. However, their
method \cite{pt} requires to detach the decay process of the
resonance $\Delta^{++} \rightarrow \pi^+p\gamma$ from the whole
radiative $\pi^+p$ process.

 The main difference between previous works and ours, is that our model
for the $\Delta^{++}$ resonance gives an amplitude for the radiative
$\pi^+p$ scattering that is gauge-invariant in the presence of finite
width effects and independent upon the parameter associated to contact
transformations. In addition, let us emphasize that the mass and width
of the $\Delta^{++}$ required to fit the total cross section data
of elastic $\pi^+p$ scattering, are consistent with the model-independent
analyses done in Ref. \cite{bernicha}.

 The dynamical model we use in this paper includes the contributions
of intermediate states with nucleons  and   $\Delta^{++,\ 0}$,\ $\rho$, \
and $\sigma$ resonances. We will assume isospin symmetry for the masses,
widths and strong couplings of the $\Delta$'s and nucleons. The effective
lagrangian densities relevant for our calculations can be found in  Ref.
\cite{Mariano}.
   Some of the couplings entering those lagrangians  can be fixed from low
energy phenomenology: $g_{\rho}^2/4\pi = 2.9$, $g_{\pi NN}^2/4\pi
= 14.3$ \cite{sigma} and the magnetic $\rho NN$ coupling
$\kappa_{\rho}=3.7$. The mass of the hypothetical $\sigma$ meson
was set to 650 MeV \cite{sigma} (see Ref. \cite{Mariano} for other
choices). The couplings $g_{\sigma}\equiv g_{\sigma \pi
\pi}g_{\sigma}NN$ and $f_{\Delta N\pi}$ are left as free
parameters to be determined from the $\pi^+p$ total cross section
data.

   In order to see how the model works in the case of elastic $\pi^+ p$
scattering, let us focus on the $\Delta^{++}$ contribution to the
$\pi^+(q)p(p) \rightarrow \pi^+(q')p(p')$ amplitude
\cite{elamiri} (letters within brackets denote four-momenta):

\begin{eqnarray}
{\cal M}(\pi^+ p \rightarrow \pi^+ p) =i\left({f_{\Delta N\pi} \over
m_{\pi}}\right)^2
\ubar{p'} q'_\mu G^{\mu\nu}(p+q) q_\nu \u{p},\nonumber
\end{eqnarray}
where $f_{\Delta N \pi}$ is the $\Delta N \pi$ coupling constant and the
$\Delta^{++}$ propagator in momentum space $ G^{\mu\nu}(p+q)$ is given in
Eq. (10) of Ref. \cite{elamiri}.
 According to the complex mass scheme, we must replace
$m_{\Delta}^2 \rightarrow
m_{\Delta}^2-im_{\Delta}\Gamma_{\Delta}$ in $ G^{\mu\nu}(p+q)$, where
$m_{\Delta}$ and $\Gamma_{\Delta}$ are the mass and width of the
$\Delta^{++}$.
As shown in \cite{elamiri}, the above amplitude is explicitly independent
of the parameter associated to contact transformations.

   Since the $\Delta^{++}$ largely dominates the elastic scattering
amplitude in  the resonance region, we expect the contributions
from other mesonic resonances and of crossed channels with
intermediate nucleon and $\Delta^0$ states to play the role of
background terms to the $\Delta^{++}$ resonance. If we add
coherently these contributions to ${\cal M}(\pi^+ p \rightarrow
\pi^+ p)$, we can fit the experimental results data for the total
cross section \cite{pedroni} of the elastic $\pi^+p$ scattering
with four free parameters ( $m_{\Delta},\ \Gamma_{\Delta},\
f_{\Delta N \pi}$ and $g_{\sigma}$) in the range $75\ {\rm MeV}
\leq T_{lab} \leq 300\ {\rm MeV}$ for energies of incident pions.
In order to compare the size of the different contributions, we
have chosen to fit the data by adding a new contribution in each
fit. The results  are shown in Table 1 and Figure 1. Let us note
that since non-resonant contributions are included at
the tree-level (they are real), and  the decay width in the
$\Delta^{++}$ propagator is taken as a constant within the complex
mass scheme, our amplitude for elastic scattering will not be unitary.
Note, however (see the appendix of Ref. \cite{Mariano}) that the terms
neglected in our approximation would be reflected in a slight increase
of the $\Delta$ decay width.

\begin{figure}[h]
\vspace{-2.5cm}
\begin{center}
    \leavevmode
   \epsfxsize = 8cm
     \epsfysize = 10cm
    \epsffile{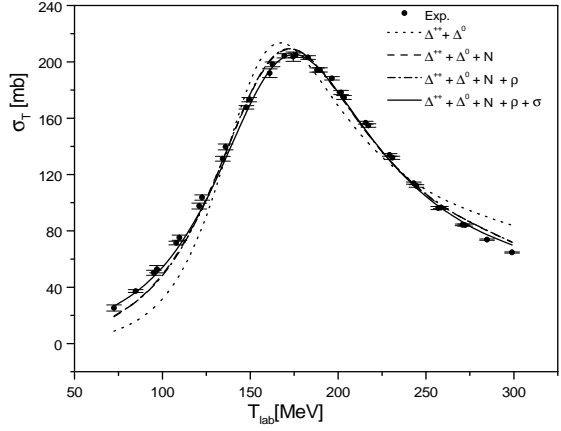}
\vspace{-3cm} \caption{ Elastic $\pi^+ p$ cross section as a
function of incident $\pi^+$ kinetic energy. } \label{fig1}
\end{center}
\end{figure}

\begin{figure}[h]
\vspace{-1.5cm}
\begin{center}
    \leavevmode
   \epsfxsize = 11cm
     \epsfysize = 11cm
    \epsffile{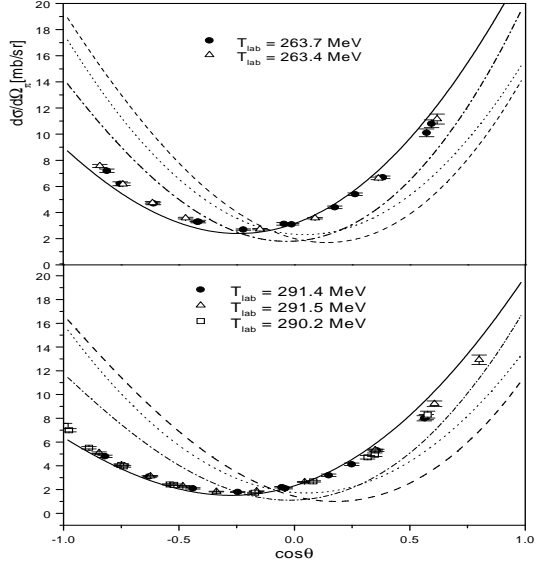}
\vspace{-3cm}

\caption{Differential cross section for elastic
$\pi^+ p$ scattering.
The curves (with same convention as in Figure 1) denote our prediction
for $T_{lab}=263.7$ (upper box) and 291.4 MeV (lower box). }
\label{fig2}
\end{center}
\end{figure}
\vspace{-0.5cm}

\begin{table}[h]
\caption{
Fits to the cross section of $\pi^+p$
scattering by including different intermediate states ($\Delta$'s,
$N,\rho, \sigma$).}
\label{table:1}
\newcommand{\m}{\hphantom{$-$}}
\newcommand{\cc}[1]{\multicolumn{1}{c}{#1}}
\renewcommand{\tabcolsep}{0.9pc} 
\renewcommand{\arraystretch}{1.2} 
\begin{tabular}{@{}llllll}
\hline
 Int. state&$f_{\Delta N \pi}^2/4\pi\ $ &$m_{\Delta}$ (MeV) &
$\Gamma_{\Delta}$ (MeV) & $g_{\sigma}/4\pi$ & $\chi^2$/dof\\
\hline $\Delta$'s & 0.281$\pm$0.001 & 1201.7$\pm$0.2 &
69.8$\pm$0.2 & -- & 121.1\\ $\Delta$'s, $N$ & 0.331$\pm$0.003 &
1208.6$\pm$0.2 & 87.5$\pm$0.3 & -- & 17.6\\ $\Delta$'s, $N, \rho$
& 0..327$\pm$0.001 & 1207.4$\pm$0.2 & 85.6$\pm$0.3 & -- & 15.6 \\
$\Delta$'s, $N,\rho, \sigma$ & 0.317$\pm$0.003 & 1211.2$\pm$0.4 &
88.2$\pm$0.4 & 1.50$\pm$0.12 & 10.5\\ \hline
\vspace{-0.75cm}
\end{tabular}
\end{table}

  Using the results of Table 1, we can
{\it predict} the angular distribution of pions in elastic $\pi^+p$
scattering. In Figure 2 we compare our prediction for the differential
cross section for $T_{lab}=263.7$ and $291.4$ MeV with the corresponding
experimental data from Refs. \cite{Bussey,Sadler,Gordeev}. It is
interesting to
observe that data in Figure 2 are well described despite the fact that
kinetic energies of incident pions in Fig. 2 lie in the upper tail of the
$\Delta^{++}$ resonance shape (see Figure 1). This is very important
because kinetic energies of incident pions in  radiative $\pi^+p$
scattering to be considered below, correspond to those particular values.

  Some interesting features are worth to be pointed out from this
analysis. First, the  agreement with data improves when the
contributions from all intermediate states \linebreak ($\Delta^{++},\ \Delta^0,\ N,\
 \rho$ and $\sigma$) are included, both for the total and the differential
cross sections (the last $\chi^2$/dof in Table 1, actually drops to 4.5
when the last three points in the cross section are excluded). Second, the
values obtained for the mass and width of
the $\Delta$'s, namely $m_{\Delta}=(1211.2 \pm 0.4)$ MeV
and $\Gamma_{\Delta}=(88.2 \pm 0.4)$ MeV,  are similar to the pole
parameters
obtained from a model-independent analysis of the same data, namely
\cite{bernicha}: $M=(1212.20 \pm 0.23)$ MeV and
$\Gamma=(97.06 \pm 0.35)$ MeV \cite{foot}. This is a non-trivial feature
given the different nature of both approaches:
in  Ref. \cite{bernicha} the amplitude for elastic $\pi^+p$ scattering
was written as a sum of a pole plus a background term as dictated by the
analytic S-matrix theory \cite{smatrix}. The above information indicates
that the contributions to the elastic scattering other that the
$\Delta^{++}$, indeed represent well the background.

  Next we focus on the determination of the $\Delta^{++}$
MDM from $\pi^+(q)p(p) \rightarrow \pi^+(q') p(p')
\gamma(\epsilon, k)$ (letters within brackets denote four-momenta and
$\epsilon$ the photon polarization).
Using the Lagrangians given in Ref.\cite{Mariano} and the complex mass
scheme to include the finite width of the $\Delta^{++}$ we obtain the
following amplitude for the resonance contribution:

\vspace{-0.5cm}
\begin{eqnarray}
{\cal M}(\pi^+p \rightarrow \pi^+p \gamma) &=&
- e \left({f_{\Delta N\pi}\over m_{\pi}}\right)^2 q'_\mu q_\nu \ubar{p'}
\left[{ \over }G^{\mu\nu}(P')
\left({ q \cdot \epsilon  \over q \cdot k} +
{p\cdot \epsilon - R(p)\cdot \epsilon \over p \cdot k}\right)
\right.\nonumber \\
&-& \left.\left({ q' \cdot \epsilon
\over q' \cdot k} +
{p'\cdot \epsilon - R(p')\cdot \epsilon \over p' \cdot k}\right)
G^{\mu\nu}(P)
+ 2 i G^{\mu\alpha}(P') \Gamma_{\alpha \beta \rho} \epsilon^\rho
G^{\beta\nu}(P) \right.\nonumber \\
&+& \left.{1 \over q \cdot
k}G^{\mu\rho}(P')F_{\rho}^{\ \nu}
 - {1 \over q' \cdot k}F^{\mu}_{\ \rho}G^{\rho\nu}(P) \right]
 u(p)\label{radiative}
\end{eqnarray}
where $F^{\rho\sigma}\equiv \epsilon^{\rho}k^{\sigma}-
\epsilon^{\sigma}k^{\rho}$, and $  R_\mu(x) \equiv {1\over 4}
[\ks,\gamma_\mu] + {\kappa_p\over 8 m_N} \{
[\ks,\gamma_\mu],\xs\}$; $e$ is the proton charge, $\kappa_p $
denotes the anomalous magnetic moment of the proton, and $P = p+q
$, $P'=p'+q'$, such that $P=P'+k$. This amplitude  is explicitly
gauge-invariant and does not depend on the parameter associated to
contact transformations \cite{elamiri}. It can also be verified
that this amplitude satisfies Low's soft photon theorem \cite{low}
as required. The electromagnetic vertex of the $\Delta^{++}$
appearing in Eq.(\ref{radiative}) is given by:
\begin{eqnarray}
\Gamma_{\alpha\beta\rho} & = &
\left( \gamma_{\rho}-\frac{i\kappa_{\Delta}}{2m_{\Delta}} \sigma_{\rho
\sigma}k^{\sigma} \right) g_{\alpha \beta} -\frac{1}{3} \gamma_{\rho}
\gamma_{\alpha} \gamma_{\beta} -\frac{1}{3} \gamma_{\alpha}g_{\beta \rho}
+ \frac{1}{3} \gamma_{\beta} g_{\alpha \rho}\ ,\nonumber
\end{eqnarray}
where $\kappa_{\Delta}$  is related to the total magnetic moment of the
$\Delta^{++}$ by $\mu_{\Delta}=2(1+\kappa_{\Delta}) (e/2m_{\Delta})$.

 The only adjustable parameter in radiative $\pi^+p$ scattering is the
$\Delta^{++}$ MDM. The contributions to this process coming from
other intermediate states ($\Delta^0$,  $\rho$, $\ N$ and
$\sigma$) can  be added to Eq. (1) in a gauge-invariant way (see
for example \cite{Mariano}). We are interested in the description
of the differential cross section
$d\sigma/d\omega_{\gamma}d\Omega_{\pi}d\Omega_{\gamma}$, as a
function of the photon energy for fixed energies of incident pions
and photon angle emission.  We have chosen to fit a subset  of
data of Ref. \cite{ucla} where photons are detected in angular
configurations as shown in Table 2. According to Ref. \cite{kpz}
one expects the differential cross section to be more sensitive to
the effects of $\mu_{\Delta}$ in this case \cite{Mariano}.
Furthermore, we chose the range of  photon energies $20\ {\rm MeV}
\leq \omega_{\gamma} \leq 100$ MeV where we expect Low's soft
photon approximation to be more reliable. Details of  the fits for
other angular configurations and its sensitivity with $\mu_\Delta$
are given elsewhere \cite{Mariano}.

   The results of the fits for $\kappa_{\Delta}$ are shown in Table 2 for
two energies of incident pions \cite{ucla}.  Figure 3 displays
the  differential cross section as a function of the photon energy for
three different geometries of photon emission (G1, G4, G7 as indicated
in Table 2) and incident pions of energy $T_{lab}=269$ MeV \cite{ucla}.
The solid line in Figure 3 corresponds to the best
fit, and the prediction when $\kappa_{\Delta}=1$ (dashed line) is shown
for comparison. Although the experimental data are rather scarce,
Fig. 3 clearly indicates  that the photon spectrum for the chosen
geometries is indeed sensitive to the effect of the $\Delta^{++}$ MDM.

\begin{figure}
\vspace{-1.5cm}
\begin{center}
    \leavevmode
   \epsfxsize = 12cm
     \epsfysize = 10cm
    \epsffile{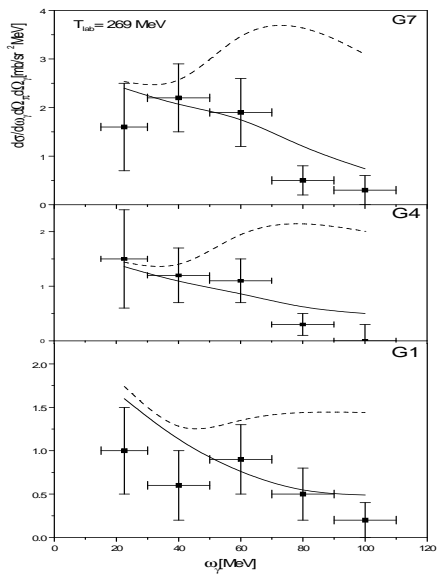}
   \end{center}
\vspace{-3.cm} \caption{ Differential cross section in radiative
$\pi^+ p$
 scattering for $T_{lab} = 269$ MeV. The  G1, G4 and G7 geometries are
defined in Table 2. The solid line corresponds to the best fit and the
dashed line  to $\kappa_{\Delta}=1$.}
\label{fig3}
\end{figure}
\vspace{-0.5cm}
\begin{table}[h]
\caption{
Fits to the differential cross section of radiative
$\pi^+p$ scattering (see text for labels).}
\newcommand{\m}{\hphantom{$-$}}
\newcommand{\cc}[1]{\multicolumn{1}{c}{#1}}
\renewcommand{\tabcolsep}{1.7pc} 
\renewcommand{\arraystretch}{1.2} 
\begin{tabular}{@{}llllll}
\hline
$T_{lab}$ (MeV)&Geometry &$\theta_{\gamma}$& $\phi_{\gamma}$ &
$\kappa_{\Delta}$ & $\chi^2$/dof  \\
\hline
& G7 & 120$^0$ & 0$^0$ & 3.27$\pm$0.76 & 1.99 \\
269& G4 & 140$^0$ & 0$^0$ & 3.01$\pm$0.67 & 2.48\\
&G1& 160$^0$ & 0$^0$ & 2.74$\pm$0.87 & 1.73 \\
\hline
&G7& 120$^0$ & 0$^0$ & 3.10$\pm$0.87 & 2.68\\
298 & G4 & 140$^0$ & 0$^0$ & 2.90$\pm$0.75 & 4.75 \\
&G1 &160$^0$ & 0$^0$ & 2.61$\pm$1.00 & 1.47\\
\hline
\end{tabular}
\end{table}

  The determinations of $\kappa_{\Delta}$ shown in
Table 2 are consistent among themselves, making meaningful to
quote a weighted average over the six different fits. If we
express the weighted average in units of nuclear magnetons we
obtain:
\begin{equation}
\mu_{\Delta} = 2(1+\kappa_{\Delta}) \frac{m_p}{m_{\Delta}} \left(
\frac{e}{2m_p} \right) =\left( 6.14 \pm 0.51 \right) \frac{e}{2 m_p}\ .
\end{equation}
The effects associated to the non-unitarity of the elastic
$\pi^+ p$ scattering amplitude have been estimated to decrease the central
value in Eq. (2) by 2\% (see appendix of Ref. \cite{Mariano}).

  This result is compatible with the prediction $\mu_{\Delta}
=5.58 (e/2m_p)$ obtained in the SU(6) quark model \cite{blp} and
with the result $\mu_{\Delta} =6.17 (e/2m_p)$ from  a recent quark
model calculation that includes non-static effects associated to
pion exchange and orbital excitations \cite{Fra}, and it is
somewhat larger than the prediction obtained from bag-model
corrections to the quark model, $\mu_{\Delta}= (4.41 \sim
4.89)(e/2m_p)$ \cite{brv}. Our results are in agreement with
previous determinations from experimental data of Refs.
\cite{wittman,pt}: 5.58$\sim$  7.53 $(e/2m_p)$ and
(5.6$\pm$2.1)$(e/2m_p)$, respectively. Our result in Eq. (2) is
larger than the one obtained from a variant of the soft-photon
approximation \cite{Lin}: $\mu_{\Delta} =(3.7 \sim 4.9)(e/2m_p)$.

  In conclusion, we have analyzed the elastic and radiative $\pi^+p$
scattering within a full dynamical model which gives amplitudes
that are gauge-invariant when finite width effects of the
$\Delta^{++}$ are introduced. These amplitudes are free of
ambiguities related to contact transformations on the spin 3/2
fields. The relevant parameters of the $\Delta^{++}$ are fixed
from the total cross section of the elastic scattering and
prediction for the differential cross section is in satisfactory
agreement with data. From a fit to the differential cross section
of the radiative process we have obtained a determination of the
$\Delta^{++}$ MDM, Eq. (2), that is in agreement with recent
predictions based on the quark model \cite{Fra}. For completeness,
let us mention that our model describes  a wider set of radiative
$\pi^+p$ data \cite{Mariano}. This is to our knowledge, the first
determination of the $\Delta^{++}$ MDM from a full dynamical model
that consistently incorporates its finite width and that is free
of ambiguities related to contact transformations

\vskip 0.5cm

{\large Acknowledgements}: The Work of A. M. was partially
supported by Conicet (Argentina). G.L.C. acknowledges partial
support from  Conacyt, under contracts 32429-E and ICM-W-8016.


\vskip 0.5cm

\begin{thebibliography}{9}
\bibitem{pdg98}
C. Caso {\it et al}, Particle Data Group, Eur. Phys. J. {\bf C3}, 1
(1998).
\bibitem{z0}
R. G. Stuart, Phys. Lett. {\bf B262}, 113 (1991); {\bf 272}, 353 (1991);
Phys. Rev. Lett. {\bf 70}, 3193 (1993); A. Sirlin, Phys. Rev. Lett. {\bf
67}, 2127 (1991); H. Veltman, Z. Phys. {\bf C62}, 35 (1994).
\bibitem{smatrix}
R. E. Peierls, {\it Proc. of the 1954 Glasgow Conf. on Nucl. and Meson
Physics}, Ed. E. H. Bellamy and R. G. Moorhouse (Pergamon Press, 1955),
p. 296; M. L\'evy, Nuovo Cim. {\bf 13}, 115 (1959);
R. Eden, P. Landshoff, D. Olive, and J. Polkinghorne,
``The Analytic S-matrix" (Cambridge University Press, Cambridge, 1966).
\bibitem{fls}
U. Baur and D. Zeppenfeld, Phys. Rev. Lett. {\bf 75}, 1002 (1995); E.
Argyres {\it et
al}, Phys. Lett. {\bf B358}, 339 (1995); M. Beuthe {\it et al} Nucl. Phys.
{\bf
B498}, 55 (1997); W. Beenakker
{\it et al}, Nucl. Phys. {\bf B500}, 255 (1997).
\bibitem{bls}
G. L\'opez Castro and G. Toledo S\'anchez, Phys. Rev. {\bf D61},
033007 (2000).
\bibitem{w}
G. L\'opez Castro, J. L. Lucio  M.,  and J. Pestieau, Mod. Phys. Lett.
{\bf A}, (1991); Int. J. Mod. Phys. {\bf A10}, (1996); A. Pilaftsis and M.
Nowakowski, Z. Phys. {\bf C60}, 121 (1993).
\bibitem{bernicha}
A. Bernicha, G. L\'opez Castro and J. Pestieau, Nucl. Phys. {\bf A597},
623 (1996).
\bibitem{sin}
A. Bosshard {\it et al}, Phys. Rev. {\bf D44}, 1962 (1991); C. A. Meyer
{\it et al}, Phys. Rev. {\bf D38}, 754 (1988).
\bibitem{ucla}
B. M. K. Nefkens {\it et al}, Phys. Rev. {\bf D18}, 3911 (1978).
\bibitem{Lin}D. Lin , M.K. Liou, and Z.M. Ding, Phys. Rev. {\bf C44},
1819 (1991).
\bibitem{low}
F. E. Low, Phys. Rev. {\bf 110}, 974 (1958).
\bibitem{elamiri}
M. El-Amiri, G. L\'opez Castro, and J. Pestieau, Nucl. Phys. {\bf A543},
673 (1992).
\bibitem{surdarshan}
K. Johnson and E.C.G. Surdarshan, Ann. Phys. {\bf 13}, 126 (1961)
\bibitem{etemadi}
L. M. Nath, B. Etemadi, and J. D. Kimel, Phys. Rev. {\bf D3}, 2153 (1971);
R. E. Behrends and C. Fronsdal, Phys. Rev. {\bf 106}, 277 (1958); J.
Ur\'\i as, Ph. D. Thesis, Universit\'e catholique de Louvain, Belgium
(1976).
\bibitem{wittman}
R. Wittman, Phys. Rev. {\bf C37}, 2075 (1988).
 \bibitem{pt}
P. Pascual and R. Tarrach, Nucl. Phys. {\bf B134}, 133 (1978).
\bibitem{Mariano}
G. L\'opez Castro and A. Mariano, nucl-th/0010045.
\bibitem{sigma}See for example: B.C. Pearce and B.K. Jennings, Nucl.
Phys.,{\bf A528},655 (1991); C. Sch\"utz, J.W. Durso, K. Holinde, and J.
Speth, Phys. Rev. {\bf C49}, 2671(1994).
\bibitem{pedroni} E. Pedroni {\it et al}, Nucl. Phys. {\bf A300},
321(1978).
\bibitem{Bussey} P.J. Bussey, {\it et al}, Nucl. Phys. {\bf B58},
363(1973).
\bibitem{Sadler} M. E. Sadler, {\it et al}, Phys. Rev. {\bf D35}, 2718
(1987).
\bibitem{Gordeev} V. A. Gordeev, {\it et al}, Nucl. Phys. {\bf A364}, 408
(1981).
\bibitem{foot} Resonance parameters defined from the pole position are
smaller than the ones obtained from a Breit-Wigner with energy-dependent
width \cite{pdg98}.
\bibitem{kpz}
V. I. Zakharov, L. A. Kondratyuk, and L. A. Ponomarev, Yad. Fiz. {\bf
8}, 783 (1968) [Sov. J. Nucl. Phys. {\bf 8}, 456 (1968)].
\bibitem{blp}
M. A. B. Beg, B. W. Lee, and A. Pais, Phys. Rev. Lett. {\bf 13}, 514
(1964).
\bibitem{brv}
G. E. Brown, M. Rho, and V. Vento, Phys. Lett. {\bf B97}, 423 (1980).
\bibitem{Fra}
J. Franklin, e-print hep-ph/0103139.

\end{thebibliography}
\end{document}